\documentclass[preprint,3p,sort&compress,12pt]{elsarticle}

\usepackage{amssymb}
\usepackage{natbib}
\usepackage{amsmath,amsthm,bm,mathrsfs}
\usepackage{color}

 \newtheoremstyle{theorem}{6pt}{6pt}{\rm}{}{\sffamily}{ }{ }{}
 \theoremstyle{theorem}

 \newtheoremstyle{algorithm}{6pt}{6pt}{\rm}{}{\sffamily}{ }{ }{}
 \theoremstyle{algorithm}

 \newtheoremstyle{lemma}{6pt}{6pt}{\rm}{}{\sffamily}{ }{ }{}
 \theoremstyle{lemma}

\newtheoremstyle{case}{6pt}{6pt}{\rm}{}{\sffamily}{. }{ }{}
 \theoremstyle{case}

 \newtheoremstyle{statement}{6pt}{6pt}{\rm}{}{\sffamily}{ }{ }{}
\theoremstyle{statement}

 \newtheoremstyle{corollary}{6pt}{6pt}{\rm}{}{\sffamily}{ }{ }{}
 \theoremstyle{corollary}

  \newtheoremstyle{definition}{6pt}{6pt}{\rm}{}{\sffamily}{ }{ }{}
 \theoremstyle{definition}

\newtheoremstyle{example}{6pt}{6pt}{\rm}{}{\sffamily}{ }{ }{}
\theoremstyle{example}

\newtheoremstyle{remark}{6pt}{6pt}{\rm}{}{\sffamily}{ }{ }{}
\theoremstyle{remark}

\newtheoremstyle{approximation}{6pt}{6pt}{\rm}{}{\sffamily}{ }{ }{}
\theoremstyle{approximation}

\newtheoremstyle{scheme}{6pt}{6pt}{\rm}{}{\sffamily}{ }{ }{}
\theoremstyle{scheme}

\newtheoremstyle{Algorithm}{6pt}{6pt}{\rm}{}{\sffamily}{ }{ }{}
\theoremstyle{Algorithm}

\newtheoremstyle{Assumption}{6pt}{6pt}{\rm}{}{\sffamily}{ }{ }{}
\theoremstyle{Assumption}

\newtheoremstyle{proposition}{6pt}{6pt}{\rm}{}{\sffamily}{ }{ }{}
\theoremstyle{proposition}

\newtheoremstyle{hypo}{6pt}{6pt}{\rm}{}{\sffamily}{ }{ }{}
 \theoremstyle{hypo}

  \newtheoremstyle{Step}{6pt}{6pt}{\rm}{}{}{ }{ }{}
 \theoremstyle{Step}

\journal{-}
\begin{document}

\begin{frontmatter}

\title{Entropy-Norm space for geometric selection of strict Nash equilibria in $n$-person games}
\author{A. B. Leoneti} 
\ead{ableoneti@usp.br}
\author{G. A. Prataviera}
\ead{prataviera@usp.br}
\address{Departamento de Administra\c{c}\~ao, FEA-RP, Universidade de S\~{a}o Paulo, 14040-905, Ribeir\~{a}o Preto, SP, Brazil}

\begin{abstract}
{Motivated by empirical evidence that individuals within group decision
making simultaneously aspire to maximize utility and avoid inequality
we propose a criterion based on the entropy-norm pair for geometric selection
of strict Nash equilibria in $n$-person games. For this, we introduce a mapping
of an $n$-person set of Nash equilibrium utilities in an Entropy-Norm space.
We suggest that the most suitable group choice is the equilibrium closest to
the largest entropy-norm pair of a rescaled Entropy-Norm space. Successive
application of this criterion allows ordering of possible Nash equilibria
in $n$-person games accounting simultaneously equality and utility of players payoffs. 
Limitations of this approach for certain exceptional cases are discussed.
In addition, the criterion proposed is applied and compared with the results
of a group decision making experiment.}
\end{abstract}

\begin{keyword}
Nash equilibrium\sep Games\sep Equlibrium selection\sep Entropy\sep
Norm


\end{keyword}

\end{frontmatter}




\section{Introduction}
\label{intro}

Nash Equilibrium is an important concept in game theory \cite{nash1,nash2,r0,r1}. In Nash equilibrium, players have assumptions about the strategies of their competitors and choose the best possible strategy to them. Other players, when acting the same, will lead to a situation where none of the participants will have the incentive to change their strategy (if they are acting in accordance to the classic Economics definition of rationality). As pointed in \cite{r1}: \textquotedblleft since each player knows the choices of others, and that is rational, their choices should be optimal given the choices of others, so by definition we have a Nash equilibrium\textquotedblright.

However, despite its interesting mathematical properties, i.e. stability, the concept of Nash equilibrium has no intrinsic or ethical meaning, having not sufficient information to be classified as a good or a bad outcome. Furthermore, in some games there may exist more than one Nash equilibrium, and some of them might involve lower payoffs. Consequently, Nash equilibrium cannot be assumed, a priori, as a satisfactory welfare result for a social choice \cite{r2}. Hence, an important question when using the concept of Nash equilibrium for analysing group decision is what would be, among many possibles Nash equilibria, the most suitable for a group choice? This is the so called equilibrium selection problem \cite{r0}. Although some theoretical approaches for solving the equilibrium selection problem has been proposed\cite{r0,r2a,r2d}, it seems that a formal approach indicating the unique ``right'' equilibrium has not been found \cite{r2c}.

In this work we explore a heuristic approach based on empirical deduction for selecting strict Nash equilibria solutions in a $n$-person game. We are proposing a more pragmatic selection criterion motivated by the empirical evidence \cite{guth1982,camerer,fehr2008,hsu2008,feng1,nr9} that individuals within a group, besides maximize their utilities, may also have fairness motivations, which can be manifested as an aversion to inequality, and so contributing to the search of a group's agreement. In fact, cooperation between individuals has been an important factor for the maintenance of human beings' lives and ongoing research efforts are been carried out in order to understand the emergence of cooperative group decisions \cite{nr9,nr1,nr2,nr3,nr4,nr5,nr6,nr7,nr8,nr10}. Nevertheless, in order to distinguish the type of frameworks available for games modelling, here we are interested in group decisions having some kind of coordination, which does not lay necessarily in the approach of cooperative games, since the term co-operate means that two or more individuals are operating as a unique entity, which is not the case of the problem addressed here, but illustrates quite well the same idea, of individuals autonomously orienting their behaviour toward a mutual benefit solution. Hence, it is proposed here that Nash equilibria in group decision making could be selected to satisfy simultaneously the principles of (i) equality; and (ii) utility.

Firstly, we adopt the concept of entropy as a measure of concentration for the players' utilities probability distribution of strict Nash equilibria solutions. From Physics \cite{reif,jaynes1} to  Information Theory \cite{r4,r5} the concept of entropy has now been used in many different contexts \cite{r4,r5,r7,r8,r10,r9,abbas,dionisio,volkov,prata}. Jaynes introduced entropy as an inference tool, the maximum entropy principle \cite{jaynes2,r7,r8,volkov}, to attribute probability for data under conditions of uncertainty, and which has been extensively used for parameter estimation in Econometrics \cite{r8}, or in decision theory as a probability functional to generated utility functions \cite{topsoe,abbas,dionisio}. In Economics, the use of entropy as a measure of inequality was introduced by Theil through the well-known Theil index \cite{r10,r9}, indicating that entropy is an interesting parameter for taking into account the principle of equality.

Secondly, in association with the entropy, we propose the use of a Euclidean norm for selecting strict Nash equilibria based on utility principle \cite{r10a}. The utility principle accounts for the typical rational behaviour of individuals assumed in Economics \cite{varian}, whose aspirations leads them to make decisions aiming higher possible utilities outcomes \cite{camerer}. However, deviation from the rational behaviour of individuals within group decision processes has been emphasized in several works \cite{guth1982,nowak2000,sanfey2003,camerer,berg1,fehr2008,hsu2008,tabibnia2008,henrich2010,rilling, feng1,bault2017,monterola,nr9}.

Thus, in order to satisfy simultaneously the principle of equality and utility, we propose a criterion for selection of strict Nash equilibrium in a n-person game based on the entropy-norm pair. For that, we introduce a mapping of Nash equilibria utilities in an Entropy-Norm space. Then, we assume that the most suitable group choice is the equilibrium closest to the largest entropy-norm pair in a rescaled Entropy-Norm Space. Successive application of this criterion permits an ordering of Nash equilibria in a $n$-person game accounting simultaneously the equality and utility of players´ pay-off. Limitations of this approach for certain exceptional cases is discussed. In addition, we apply the criterion proposed to analyse the results of a game decision experiment.

The article is organized as follows: in Section (\ref{sec2}) we introduce Entropy-Norm Space and the criterion for geometric selection of the most suitable set of Nash equilibria for a group decision. In Section (\ref{sec3}) we apply and discuss the criterion proposed to a group decision experiment. Finally, in Section (\ref{sec4}) we present the conclusions.

\section{Entropy-Norm Space and equilibrium selection}\label{sec2}

\label{sec:1}
Let us consider a strategic $n$-person game defined by the tuple $\langle \bold{N}, A, \succ_i \rangle $, where $\bold{N}$ is the set of $n>1$ players (decision makers), \textit{A} is the set of \textit{m} actions (alternatives), and $\succ_i $ is the preference set over \textit{A} for each player $i \in \bold{N}$. If any, to each strict Nash equilibrium found we can  associate a vector $u=(u_{1}, u_{2}, \cdots , u_{n})$ with components corresponding to the utilities of the player $1$, player $2$, $\cdots$, and player $n$, respectively. To each component $u_{i}$ of a given strict Nash equilibrium vector we can associate a probability $p(u_{i}) $ through its respective utility proportion, that is, 
\begin{equation}
p\left( u_{i}\right)=\frac{u_{i}}{\sum_{i=1}^{n} u_{i}},\hspace{1cm} i=1,2,. . ., n, 
\label{prob}
\end{equation}
such that $\sum_{i=1}^{n}p\left(u_{i}\right)=1$. Then, the entropy $S_{u}$ associated with the probability distribution of an equilibrium vector $u$ is given by \cite{r4,r5}
\begin{equation}
S_{u}=-\sum_{i=1}^{n}p\left(u_{i}\right)\ln p\left(u_{i}\right),
\label{entro}
\end{equation}
with $0\leq S_{u} \leq \ln n$, where $S_{u}=0$ corresponds to only one of the $p\left( u_{i}\right)=1$, while the maximum value $S_{u}=\ln n$ is attained for a uniform probability distribution, corresponding to an equilibrium vector in which all players have the same utilities values. Hence, entropy is useful to quantify inequality in the distribution of utilities of a given vector $u$. 

Complementary, in order to take into account the utility principle we need introduce a measure of intensity. For that we associate with an strict Nash equilibrium vector $u$ its Euclidean norm
\begin{equation}
\lVert u\rVert =\sqrt{u_{1}^2+u_{2}^2+u_{3}^2+\cdots +u_{n}^2}.
\end{equation}

Thus, to each element of a discrete set $U=\{u^{(1)},u^{(2)},\cdots ,u^{(k)}\}$ of $k$ strict Nash equilibrium utility vectors we can associate their entropy and norm, and each utility vector can be mapped in a point of a bi-dimensional Entropy-Norm Space. For this it is convenient to settle both entropy and norm measures in scales with same range by using the ordering preserving map $\rho: R \Rightarrow [0,1]$ given by
\begin{equation}
\label{escala}
\rho \left(x\right)=\frac{x-x_{min}}{x_{max}-x_{min}},
\end{equation}
where $x$ is the value to be rescaled of the entropy and norm associated with each equilibrium vector $u^{(j)}$, and $x_{min}$ and $x_{max}$ are the minimum and maximum values of entropy and norm in the set $U$, respectively. 

Now, we introduce the Entropy-Norm Space as an orthogonal bidimensional space, which is composed by a pair generated by the rescaled entropy and norm measures for a set of strict Nash equilibrium vectors, where an element of this space is defined as a pair $\left( \rho \left(S_{u}\right),\rho\left(\lVert u \rVert \right)\right)$. For a $n$-person game we propose the solutions to be 
geometric selected from the Entropy-Norm Space considering the largest values of 
equality and utility based on entropy and norm. In the rescaled Entropy-Norm Space the 
maximum of equality and utility corresponds to the the point $(1,1)$. Then, assuming the 
existence of a set $U=\{u^{(1)},u^{(2)},\cdots ,u^{(k)}\}$ with $k$ strict Nash equilibria, 
the equilibrium to be selected is any strict Nash equilibrium vector $u$ corresponding to a 
point at Entropy-Norm Space satisfying the following criterion:

\vspace{.5cm}
\textit{Let $\{d_{u^{(1)}}, d_{u^{(2)}},\cdots ,d_{u^{(k)}}\}$ be the set of Euclidean distances of each equilibrium point $\left(\rho \left(S_{u^{(j)}}\right),\rho\left(\parallel u^{(j)}\parallel \right)\right)$ to the point $(1,1)$ at the Entropy-Norm Space, where
\begin{equation}
d_{u^{(j)}}=\sqrt{[1-\rho(\lVert u^{(j)} \rVert)]^2+[1-\rho(S_{u^{(j)}})]^2} \hspace{0.5cm} j=1,\cdots,k.
\end{equation}
Then,
\newline (a) select the equilibrium vectors set $U^{*}=\{u_{1}^{*}, u_{2}^{*},\cdots,u_{j\leq k}^{*}\}$ satisfying 
\begin{equation}
d_{u^{*}}=\text{min}\{d_{u^{(1)}}, d_{u^{(2)}},\cdots,d_{u^{(k)}}\}.   
\end{equation}
(b) from the set $U^{*}$ select the equilibrium vectors subset whose map in the Entropy-Norm Space is closer to the line $\rho (S_{u}) =\rho (\lVert u \rVert)$.}
\vspace{.5cm}

The criterion allows the selection of the Nash equilibrium geometrically taking simultaneously into account the equality and utility principles. Moreover, it follows that equilibrium states in the Entropy-Norm Space with same distance to the point $(1,1)$ are equivalent. This means that the solution may not be unique since all equilibrium states mapped at the same point in the Entropy-Norm Space are strictly equivalent, and therefore indistinguishable based on the equality-utility principle. Thus, the condition (b) guarantees a more parsimonious choice between equivalent points with different coordinates at the Entropy-Norm Space. An ambiguity occurs only if there exists equivalent points localized symmetrically to the line $\rho (S_{u}) =\rho (\lVert u \rVert )$. Then the individuals should consider the \text{tradeoff} between equality and utility, or some dynamical aspect of games needs to be included in order to introduce a symmetry breaking for these exceptional cases, i.e., a random choice. Notwithstanding the mentioned exceptional cases, the successive application of the criterion gives an ordering for the equilibria found.

\section{Application}\label{sec3}

To apply the entropy-norm criterion for selection of Nash equilibrium in a group decision experiment we consider empirical data adapted from Ref. \cite{r11}. The decision problem is related to the choice of a travel destination to be held in a group of five undergraduate students from the Business Management course of the School of Economics, Business Administration and Accounting at the University of S\~{a}o Paulo, at Ribeir\~ao Preto, Brazil. The information regarding five possible travel destinations  (named from A to E) was given by a performance matrix obtained from eight criteria based on five real travel destinations package offered by Brazilian tourism agencies. The players utilities for strategic choice were generated by the utility function proposed in \cite{r12} and detailed in \ref{apendice}. Table \ref{tabela} shows the strict Nash equilibria obtained from the students preferences, stressing the fact that there was no equilibrium involving the alternative A.

\begin{table}
	\caption{Pure Nash equilibria and respective students utilities for the travel destination choice group decision experiment. Utility values were generated by Eq. \ref{util} in \ref{apendice}. NE and Std stand for Nash Equilibrium and Student, respectively. Adapted from Ref. \cite{r11}.}
	\begin{tabular}{ll}
		\hline\noalign{\smallskip}
		Equilibria & Utilities  \\
		\noalign{\smallskip}\hline\noalign{\smallskip}
		\begin{tabular}{llllll}
			& Std1 & Std2 & Std3 & Std4 & Std5  \\
			\noalign{\smallskip}\hline\noalign{\smallskip}
			NE1 & B & B & B & B & B \\  
			NE2 & C & C & C & C & C \\  
			NE3 & D & D & D & D & D \\  
			NE4 & E & E & E & E & E   
		\end{tabular} &
		\begin{tabular}{lllll}
			Std1 & Std2 & Std3 & Std4 & Std5  \\
			\noalign{\smallskip}\hline\noalign{\smallskip}
			0.100 &	0.201 &	0.129 &	0.167 &	0.157 \\
			0.103 &	0.081 &	0.106 &	0.245 &	0.482 \\
			0.396 &	0.507 &	0.254 &	0.179 &	0.185 \\
			0.206 &	0.391 &	0.420 &	0.185 &	0.086
		\end{tabular} \\
		\noalign{\smallskip}\hline
	\end{tabular}
	\label{tabela}
\end{table}

As described in \cite{r11}, although most decision makers defended alternative D (NE3), a student (Student 1) firmly opposed it, leading the group to choose alternative C (NE2). This can be explained when it is assumed that players in a game with a single interaction usually will try to maximize their profits. However, from the point of group decision making to the point of implementation there might be changes in the commitment of the players. Therefore, the concern with the equality criterion in the same magnitude that the utility criterion is convenient to avoid break of contract. In the experiment, a Likert-type scale was used for measuring the players satisfaction relatively to their agreement. It was verified that the group reached a low average for both justice sense (equality principle) and satisfaction (utility principle).

\begin{figure}[!ht]
	\begin{center}
		\includegraphics[width=\linewidth]{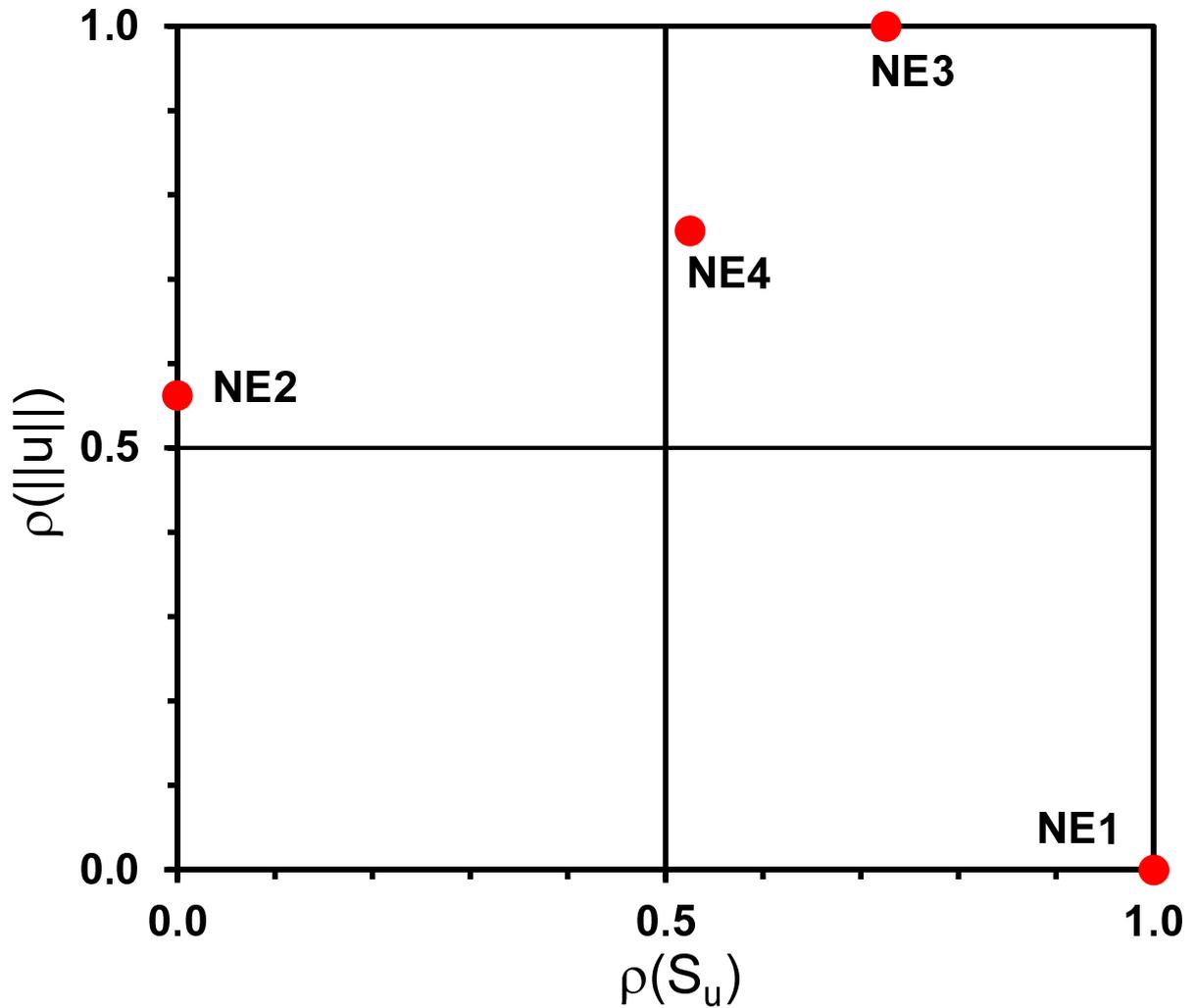}
	\end{center}
	\caption{Entropy-Norm Space Nash equilibria for the travel destination choice group decision experiment: Red dots correspond to the pair $\left( \rho \left(S_{u}\right),\rho\left(\lVert u \rVert \right)\right)$ obtained from each set NE of utilities in Table \ref{tabela}.}
	\label{fig1}
\end{figure}

Now we compare the results of ref. \cite{r11} with that given by the entropy-norm criterion. The corresponding entropy-norm pair scaled according to Eq. \ref{escala} for each equilibrium was obtained from its respective set of utilities in Table \ref{tabela}. Fig. (\ref{fig1}) shows the Entropy-Norm Space where the red points correspond to the Nash equilibria found in that game. In fact, the best choice for the group based on the entropy-norm selection criterion is unique and corresponds to the equilibrium NE3 (alternative D), followed by equilibria NE4 (alternative E). Also, from Fig. (\ref{fig1}) we verify that the Nash equilibrium solution corresponding to the point NE1 (alternative B) could lead the agents to a situation of mutual losses. While this equilibrium has the highest equality among the players, it has the worst utility, which make it an undesirable equilibrium. Furthermore, the equilibrium NE2 (alternative C) chosen by the group could not lead to high level of group justice sense since it has the lower entropy, meaning that the gains are extremely unbalanced. Actually, in \cite{r11} the authors verified that Student 1 showed high satisfaction and sense of justice regarding the group decision, while others students not, which was correctly valued in the Entropy-Norm space. Although the behaviour of Player 1 is typical of a
rational decision-maker, since it was justified that alternative D was a destination already known, in a group context this behaviour may difficult the decision implementation and create future retaliations by unsatisfied group members. Thus, contrasting with the group decision in the experiment, the entropy-norm criterion provides a more equilibrated choice of Nash equilibria satisfying simultaneously the equality and utility principles. Furthermore, it also gives evidence when the group seems to make decisions not jointly following the equality and utility principles, as occurred in the experiment, which might cause implementation problems.

\section{Conclusion}\label{sec4}

In this work we have considered the problem of selection of strict Nash equilibrium
in a $n$-person game. We followed a heuristic approach based on empirical
evidence that individuals within group decision process may prioritize
both utility and equality principles. For such, we introduced the mapping of
Nash equilibria in an Entropy-Norm space. Then,
we proposed to select from the set of Nash equilibria the ones closest to the
highest entropy-norm pair at a rescaled Entropy-Norm space. If not unique, the
equilibria selected are at least equivalent. We also have pointed the limitations
of our approach for certain exceptional cases. In addition, the entropy-norm
criterion was applied and compared with the results of a group decision
making experiment, showing that the entropy-norm criterion provides
more suitable choices that take into accounts both satisfaction and sense of
justice of players. Finally, by reversing the reasoning, the entropy-norm criterion
can be useful in empirical studies for testing if group decisions are carried
out by simultaneously satisfying inequality aversion and utility maximization
principles. Since our proposal is entirely based on the utility functions we expect it can be extended to other situations, such as those in the growing field of quantum game theory \cite{quantumgame1,quantumgame2,quantumgame3,quantumgame4}. Thus we believe that this work can contribute to the field of decision analysis and stimulate further theoretical or empirical studies on Nash equilibrium selection problem.

\vspace{0.5cm}
The authors thanks Conselho Nacional de Desenvolvimento  Cient\'{i}fico e Tecnol\'{o}gico (CNPq), Brazil, Grant Number $458511/2014\textendash 5$.

\section*{References}
\bibliographystyle{elsarticle-num} 
\bibliography{references}


%
%

\appendix

\section{Utility function}
\label{apendice}

We are using for modelling the non-cooperative game $\langle \bold{N}, A, \succ_i \rangle $ the function $\pi: R_{+}^{c\times n} \Rightarrow [0,1]$, proposed by \cite{r12}, for the numeric representation of the set of preferences $\succ_i$. This utility function shows the payoffs for a decision game among decision makers that has three strategies: (I) maintain the initial choice; (II) choose the alternative proposed by a counterpart; and (III) choose a different alternative from the alternatives proposed by a counterpart. For a game with \textit{n} players the joint utility function is given by 
\begin{equation}
\pi_{i} (x,Y)=\varphi (x,IA)\prod_{\scriptsize \begin{array}{cr}
	j=1   \\ 
	j\neq i  
	\end{array}}^{n}\varphi (x,y_{j})\varphi (y_{j},IA)
\label{util}
\end{equation}
where $\pi (x,Y)$ defines, for a determined player, the payoff for all strategies (I, II or III) for an alternative ${\mathbf x}$ when trading it with another set of alternatives $Y(y_{j})$ proposed by all other players, $IA$ is the ideal alternative (the alternative composed with the best values of each criteria), $\varphi (x,IA) , \varphi(y_{j},IA)$ and  $\varphi (x,y_{j})$ are given by the pairwise comparison function $\varphi: R_{+}^{c} \Rightarrow [0,1]$, according to
\begin{equation}
\varphi (x,y)=\left[ \frac{\alpha_{x,y}}{\| y \|}\right] ^{\delta}\cos(\theta_{x,y}),\hspace{1cm} \delta=
\begin{cases}
1, if \alpha_{x,y} \leq \| y \|  \\ 
-1, otherwise  
\end{cases}
\end{equation}
where $\alpha_{x,y}=\parallel x\parallel\cos\theta_{x,y} $ is the scalar projection of the vector $\mathbf{x}$ on the vector $\mathbf{y}$ , $\theta_{x,y}$ is the angle between the two vectors, and $\parallel\mathbf{y}\parallel=\sqrt{y_{1}^2+y_{2}^2+\cdots +y_{c}^2}$ is the norm of the respective vector. The range of $\varphi$ varies between 0 and 1 (due to the conditional $\delta$), meaning that the closer it is to $1$ the more similar are the alternatives. The use of this utility function generates the payoff tables for all players, which estimates a utility measure for every possible strategy in the set of actions. Mathematically, if one of the terms (pairwise comparison function) of the utility function is close to zero (low similarity between any pair of alternative), then $\pi_{i}(x,Y)$ tend to zero, which means that only similar alternatives closed to IA are going to be considered in what is called "acceptable space" of the game.

\end{document}